\documentclass[12pt,draftcls,onecolumn]{IEEEtran}
\usepackage{amsmath,amssymb,epsfig,setspace}
\newcommand{\figref}[1]{Fig.~\ref{fig:#1}}
\global\def\putFrag#1#2#3#4{
\begin{figure}[bp]
\begin{center}
#4
\epsfxsize=#3in
\epsfbox{#1.eps}
\end{center}
\caption{\small{#2}}
\label{fig:#1}
\end{figure}
}

\newcommand{\Prob}{\textrm{Pr}}
\newcommand{\beq}{\begin{equation}}
\newcommand{\enq}{\end{equation}}
\newcommand{\beqa}{\begin{eqnarray}}
\newcommand{\enqa}{\end{eqnarray}}
\newcommand{\ex}{{\mathbb E}}
\newcommand{\no}{\nonumber}

\newtheorem{theorem}{Theorem}
\newtheorem{proposition}[theorem]{Proposition}

\begin{document}
\newcommand{\pderiv}[2]{\frac{\partial #1}{\partial #2}}

\title{ARQ Diversity in Fading Random Access Channels}

\author{Young-Han Nam, Praveen Kumar Gopala and
Hesham El Gamal 
}

\pagestyle{plain}

        \textwidth 6.5 in
        \oddsidemargin 0 in
        \evensidemargin  0 in
        \textheight 9.25 in
        \topmargin -0.65 in

\date{}

\maketitle 

\begin{abstract}
A cross-layer optimization approach is adopted for the design of
symmetric random access wireless systems. Instead of the
traditional collision model, a more realistic physical layer model
is considered. Based on this model, an Incremental
Redundancy Automatic Repeat reQuest (IR-ARQ) scheme, tailored to
{\bf jointly} combat the effects of collisions, multi-path fading,
and additive noise, is developed. The Diversity-Multiplexing-Delay
tradeoff (DMDT) of the proposed scheme is analyzed for
fully-loaded queues, and compared with that of Gallager tree
algorithm for collision resolution and the network-assisted
diversity multiple access (NDMA) protocol of Tsatsanis~{\em et
al.}. The fully-loaded queue model is then replaced by one with
random arrivals, under which these protocols are compared in terms
of the stability region, average delay and diversity gain.
Overall, our analytical and numerical results establish the
superiority of the proposed IR-ARQ scheme and reveal some
important insights. For example, it turns out that the performance
is optimized, for a given total throughput, by maximizing the
probability that a certain user sends a new packet and
minimizing the transmission rate employed by each user.
\end{abstract}

\pagenumbering{arabic}

\section{Background} \label{sec:intro}
We consider a random access system with symmetric users who
compete to communicate with a common receiver, or a base station
(BS). Traditional approaches for analyzing such systems use the
simplified collision model (\cite{BG:92} and references therein),
which assumes that a message is received error-free by the BS {\bf
if and only if} a single user transmits. Under this model, several
protocols, which attempt to avoid collisions, have been proposed
in the literature, for example, Gallager tree algorithm (GTA) 
\cite{G:78}. The collision model, however,
does not adequately capture some important characteristics of the
wireless channel, e.g., multi-path fading, and ignores certain
physical layer (PHY) properties like multi-packet reception
(MPR)~\cite{NMT:05}. Recently, several researchers have
started to focus on cross-layer optimization approaches which
leverage the wireless medium to improve the performance of random
access systems. For example, Naware {\em et al.} \cite{NMT:05}
analyzed the stability and average delay of slotted-ALOHA based
random access channels with MPR at the BS. This analysis, however,
abstracts out the physical layer parameters by using a very
simplified model for MPR probabilities. Another example is
\cite{TZB:00} where Tsatsanis {\em et al.} propose the
Network-assisted Diversity Multiple Access (NDMA) protocol, which
uses a {\bf repetition based} Automatic Repeat reQuest (ARQ)
approach for collision resolution. As argued in the sequel, this
protocol suffers from a significant loss in throughput resulting
from repetition coding. In \cite{CT:01}, Caire {\em et al.}
analyzed the throughput of incremental redundancy (IR)-ARQ for the
Gaussian collision channel with fully-loaded\footnote{Each queue
has infinite packets for transmission.} queues and single-user
decoders at the base station. By adopting the fully-loaded queuing
model, this work ignores the stability issues that arise in
practical random access systems with random arrivals. Moreover,
the single-user decoders used in this work are sub-optimal and
result in considerable throughput losses. To overcome the
limitations of these previous works, we adopt a more realistic
model for the physical layer, and develop a variant of IR-ARQ
protocols optimized for random access over wireless channels.

\section{ARQ Random Access} \label{sec:sys_model}
In this section, we introduce our system model and briefly review
two existing random access schemes; namely, the GTA and NDMA
protocols. To the best of the authors' knowledge, these two
approaches represent the state of the art in the design of random
access protocols. We then present an
IR-ARQ random access protocol that overcomes the limitations of
these protocols.

\subsection{System Model}
We consider a $K$-user symmetric random access channel with $M$
antennas at each user and $N$ antennas at the receiver (base
station). We assume that all the channels are independent and
experience Rayleigh-flat and long-term static block fading where
the channel coefficients remain constant during one collision
resolution (CR) epoch and change independently from one epoch to
the next (a CR epoch will be defined rigorously in the next
section). The channel coefficients are assumed to be perfectly
known to the BS, but unknown to the random access users. We
consider individual power constraints on the users, and denote the
average received signal-to-noise ratio (SNR) of each user by
$\rho$. In our model, time is slotted and a slot is composed of
$T$ channel uses. In order to control the number of users
colliding in any particular slot, each user selects a slot for
transmitting a new packet according to the probability-$p_t$ rule:
in every slot, each user with a non-empty queue transmits a packet
with probability $p_t$ and does not transmit with probability
$1-p_t$, where $0 < p_t \le 1$. We assume that the BS can
perfectly identify the set of active users (by assigning a
different control channel to each user). We initially assume
fully-loaded queues in Section~\ref{sec:dmdt}, and then relax this
assumption and consider a queuing system with random arrivals in
Section~\ref{sec:random}.

\subsection{Gallager Tree Algorithm (GTA)}
This algorithm was proposed by Gallager \cite{G:78} for the random
access channel under the simplified collision model. The extension
of this algorithm to our channel model mainly includes the
probability-$p_t$ rule and an explicit assumption that the base
station does not try to decode in the case of a collision. We
describe the extended GTA as follows. The traffic in the channel
is interpreted as a flow of collision resolution (CR) epochs. At
the beginning of a CR epoch, each user uses the probability-$p_t$
rule to decide whether it should (or should not) transmit in that
epoch. If none of the users choose to transmit, the slot remains
idle and a new CR epoch starts from the following slot. If only
one user chooses to transmit, then the  message is assumed to be
successfully decoded at the BS, and a new CR epoch begins from the
following slot. But when a collision occurs, i.e., more than one
user chooses to transmit in the current slot, the system enters
into a CR mode, and only the users that participated in the
collision at the beginning of a CR epoch are allowed to transmit
until the end of that CR epoch. The colliding users are randomly
split into two different groups according to a fair random split,
wherein each user has an equal probability of joining either of
the groups. A CR epoch is finished when all the users who have
initiated it and not been excluded (or \emph{pruned}) by the tree
algorithm, obtain a slot to transmit their packets without
collisions. We omit the detailed description of the algorithm for
brevity and refer interested readers to \cite{G:78,MP:93}.

\subsection{Orthogonal Network-Assisted Diversity Multiple Access (O-NDMA)}
\label{subsec:ndma} The NDMA protocol was proposed by Tsatsanis
{\em et al.} \cite{TZB:00} and relies on the use of time diversity
through a repetition ARQ scheme to resolve collisions between
users. At the beginning of each CR epoch, the transmission of each
user will be determined by the probability-$p_t$ rule as in the
GTA protocol. If none or only a single user choose to transmit,
then the next CR epoch starts from the following slot as before.
However, when $k$ ($\ge 2$) users transmit, then all those users
repeat their transmissions in the next $(k-1)$ slots. At the end
of $k$ slots, the BS is assumed to be able to reliably decode the
$k$ packets, and a new CR epoch begins from the next slot. 
On the other hand, in~\cite{ZST:02}, Zhang {\em et al.} 
proposed a new variant of NDMA which does not rely on time diversity to
resolve/detect collisions. This variant, named B-NDMA, relies on a
blind signal separation method utilizing a Vandermonde mixing
matrix constructed via specially designed user retransmissions. In
B-NDMA, the detection and resolution of a $k$-user collision
require $(k+1)$ slots. However, in this paper, we assume the use
of separate control channels for collision detection; which allows
for a slightly more efficient version of the B-NDMA protocol, named
orthogonal NDMA (O-NDMA), which requires only $k$ slots to resolve
a $k$-user collision, without relying on temporal diversity. The
behavior of users in O-NDMA is the same as that in NDMA, with the
only difference that in case of a $k$-user collision, user $i$
transmits its symbols scaled by $(w_i)^{\ell}=(e^{\frac{j2\pi i}{k}})^{\ell}$, 
where $i=1, \cdots, k$ and $j = \sqrt{-1}$, in the $\ell$-th
slot after the initial collision. At the
end of the $k^{th}$ slot, the BS utilizes the orthogonal structure
constructed by $w_i$'s to decompose the joint decoding problem into 
$k$ single-user problems.
For example, suppose that user 1 and user 2 have collided ($k=2$), and
user $i$'s codeword is $\mathbf{x}_i$, for $i=1,2$. Then, the BS
coordinates the users so that user 1 repeats $\mathbf{x}_1$
whereas user 2 transmits $-\mathbf{x}_2$, in the slot following
the collision. To decode user 1, the BS calculates the sum of the
received vectors in the two slots, while to decode user 2, it
takes the difference (i.e., matched filtering). This way, the
multi-user interference is removed, and single-user decoders can
be utilized to recover both packets. It is worth noting that
O-NDMA requires symbol-level synchronization to
facilitate the interference cancellation described above.
Hence, our results for O-NDMA can be interpreted as optimistic
upper bounds on the performance of repetition based random access
protocols.

However, O-NDMA is still sub-optimal for two reasons. First, the
BS might be able to decode\footnote{Multiple messages can
be jointly decoded in a single transmission block, with an
arbitrary small error probability, if a rate-tuple lies within the
capacity region of the channel and a sufficiently large block
length is used \cite{G:68}.} the messages of $k$ colliding users
in less than $k$ time slots. Conversely, it is also possible that
$k$ time slots are insufficient for the successful decoding of the
$k$ packets. Thus, such a static strategy may result in a
throughput loss. Second, O-NDMA is essentially
{\bf a repetition based} collision resolution mechanism. Although
this results in a low-complexity decoder at the BS, the throughput
performance is highly sub-optimal, as shown rigorously in the
sequel. A significant improvement in the throughput can be
achieved by allowing for IR transmissions from the colliding users
within the CR epoch, and using joint decoding, across ARQ rounds
and users, at the base-station (as discussed next).

\subsection{IR-ARQ Random Access}
To overcome the disadvantages of the existing protocols, we
propose a new IR-ARQ random access protocol operating as follows.
Each user encodes an information message (packet) of $B_T$ bits
using a codebook of length-$LT$ codewords, where $L$ is an integer
denoting a deadline constraint on the transmission delay (i.e., a
constraint on the maximum number of allowed ARQ rounds). Codewords
are divided into $L$ sub-blocks of length $T$. At the beginning of
each CR epoch, the users choose to transmit or not based on the
probability-$p_t$ rule as before. Once a user chooses to transmit
in a particular slot, it transmits its first $T$ symbols during
that slot. On receiving signals at the end of a slot, the BS uses
a joint decoder that decodes the received observations both across
users and ARQ rounds. If the receiver successfully decodes {\bf
all} the transmitted messages, it feeds back an ACK bit;
otherwise, it returns a NACK signal. On receiving an ACK, the CR
epoch is terminated and a new CR epoch starts from the next slot.
Thus a CR epoch can be defined as the time between two successive
ACKs from the receiver (we observe that this definition requires
the BS to return an ACK message after an idle slot). On the other
hand, if a NACK is received, each colliding user sends its second
sub-block of $T$ codeword symbols in the next slot, while all the
other users remain silent. The ACK/NACK rule applies in a similar
manner, until the $L^{th}$ slot is reached (after $(L-1)$
consecutive NACKs). In this case, the receiver sends an ACK
regardless of its decoding result. If a certain message is decoded
after $\ell$ ARQ rounds, the effective coding rate for the
corresponding user becomes $R/\ell$ bits per channel use (BPCU),
where $R=(B_T/T)$ denotes the rate computed assuming only one
transmission round. Finally, we note that, unlike the O-NDMA, the
IR-ARQ protocol requires only slot-synchronization.

\section{The Diversity-Multiplexing-Delay Tradeoff (DMDT)} \label{sec:dmdt}

In this section, we analyze the DMDT of the proposed IR-ARQ
protocol and contrast it with our two benchmark protocols under
the assumption of fully-loaded queues. The ``fully-loaded''
assumption allows for analyzing the maximum achievable throughput
without focusing on the stability and average delay issues, for
the moment.

\subsection{Definitions}
We borrow the notion of DMDT from \cite{ECD:04}. This notion is a
generalization of the Zheng-Tse diversity-multiplexing tradeoff
(DMT) which characterizes the fundamental limits of fading
channels in the high SNR regime~\cite{ZT:02}. The delay here
refers to the {\bf maximum transmission delay} corresponding to
our upper bound $L$ on the number of ARQ rounds (including the
first one). In particular, we consider a family of ARQ protocols
where the size of the information messages $B_T(\rho)$ depends on
the operating SNR $\rho$. These protocols are based on a family of
space time-codes $\{C_\rho\}$ with a first round rate of $R(\rho)
= B_T(\rho)/T$ and an overall block length $TL$. For this family
of protocols, we define the first round multiplexing gain $r$ and
the {\bf effective} ARQ multiplexing gain $r_e$ as
\beq
r ~=~ \lim_{\rho \rightarrow \infty} ~\frac{R(\rho)}{\log_2 \rho} ~ \qquad
\mbox{and} \qquad
r_e ~\triangleq~ \lim_{\rho \rightarrow \infty}~
\frac{\eta_{FL}(\rho)}{ \log_2 \rho}.
\enq
Here $\eta_{FL}(\rho)$ is the average throughput of the ARQ
protocol in the random access channel with Fully-Loaded (FL)
queues, i.e., \beq \eta_{FL}(\rho) ~=~ \lim_{s \rightarrow
\infty}~ \frac{b(s)}{sT}, \enq where $s$ is the slot index and
$b(s)$ is the total number of message bits transmitted up to slot
$s$. Note that the message bits received in error at the BS are
also counted in $b(s)$. The {\bf effective} ARQ diversity gain is
defined as \beq d ~=~ - \lim_{\rho \to \infty} ~\frac{\log_2 P_e
(\rho)}{\log_2 \rho}, \label{eqn:d_def} \enq where $P_e(\rho)$ is
the system error probability, which is defined as the probability
that at least one of the messages is not correctly decoded by the
BS. In the symmetric random access channel, the diversity gain
obtained from \eqref{eqn:d_def} is the same as the diversity gain
of an individual user, since \beq
P_{e^{(i)}}(\rho) \le P_e(\rho) \le \sum_{j=1}^{K}
P_{e^{(j)}}(\rho), \quad \forall i \in \{1, \cdots, K\} ~, \enq
\cite{TVZ:04} 
where $P_{e^{(i)}}(\rho)$ is the error probability of the $i^{th}$
user. In summary, the DMDT of a certain protocol characterizes the
set of achievable tuples $(d,r_e,L)$ (here, we observe that our
results are information theoretic in the sense that we assume the
use of random Gaussian codebooks~\cite{ZT:02}).

In our analysis, we will make use of the results of Tse {\em et al.}
on the diversity-multiplexing tradeoff of {\bf
coordinated} multiple access channels
\cite{TVZ:04}, where the access mechanism is controlled by
the base-station. In the sequel, we denote the diversity gain of
the coordinated multiple access channel with $k$ users as
$d_k^{MAC}(r)$, which is given by \beq \label{eqn:dmac}
d_{k}^{MAC}(r) ~=~ \left\{
\begin{array}{ll} d^{M,N}(r), & r \le \min\{M,
\frac{N}{k+1}\} \\ d^{kM,N}(kr), & r \ge \min\{M,\frac{N}{k+1}\}
\end{array} \right. ~,
\enq where $d^{M,N}(r)$ is the diversity gain of the
point-to-point channel with $M$ transmit and $N$ receive antennas,
and multiplexing gain $r$, as given in \cite{ZT:02}.

In the ARQ setting, we denote the event that a NACK is transmitted
in the $\ell^{th}$ ARQ round, when $k$ users are transmitting
simultaneously, by $\bar{\mathcal{A}}_{k}(\ell)$, for $\ell = 1,
\cdots, L-1$, and the error event in the $L^{th}$ round by
$\bar{\mathcal{A}}_{k}(L)$. We also denote the complement of
$\bar{\mathcal{A}}_{k}(\ell)$ by $\mathcal{A}_{k}(\ell)$. We
define $\alpha_k(\ell) ~\triangleq~ \Prob
\left( \bar{\mathcal{A}}_{k}(1), \cdots,
\bar{\mathcal{A}}_{k}(\ell-1), \mathcal{A}_{k}(\ell) \right)$
and $\beta_k(\ell) ~\triangleq~ \Prob
\left(\bar{\mathcal{A}}_{k}(1),\cdots, \bar{\mathcal{A}}_{k}(\ell)
\right)$ for $\ell=1, \cdots, L$, where, by definition, we
let $\beta_k(0) = 1$, for $k=1, \cdots, K$. Note that
$\alpha_k(\ell)$ is the probability that the length of a CR epoch
is $\ell$ (slots), given that $k$ users have collided initially.
Following the approach of \cite{TZB:00}, we classify the CR epochs
from the viewpoint of a particular user (say user 1) into either
{\em relevant} or {\em irrelevant} epochs, depending on whether a
packet of that particular user is being transmitted in that CR
epoch or not. The lengths of the relevant and the irrelevant
epochs of user 1 are random variables, which are denoted by $U$
and $V$, respectively. For notational convenience, we denote the
pmf of a Bernoulli random variable with population $K$ and
probability of success $p$ by, $\mathcal{B}(K,k,p) ~\triangleq ~
{K \choose k} ~p^{k} (1-p)^{K-k} $.

\subsection{Main Results}
First, we characterize the DMT for GTA (note that we do not have a
deadline in this protocol, and hence, no limit on the maximum
transmission delay).

\begin{proposition} \label{thm:dmdt_gta}
The DMT for GTA with a given $p_t \in (0,1]$ is \beq
\label{eqn:thm_tree1} d^{GTA}(r_e) ~=~ d_1^{MAC}\left(
\frac{\sum_{k=0}^K \mathcal{B}(K,k,p_t) \mathcal{X}_k}
{\sum_{k=0}^K \mathcal{B}(K,k,p_t) J_k} ~r_e \right), \enq where
$\mathcal{X}_k$ and $J_k$ can be found by the following recursions:
\beq
\mathcal{X}_k ~=~ 1 + \mathcal{B}(k,0,0.5)\mathcal{X}_k + \mathcal{B}(k,1,0.5)(1+
\mathcal{X}_{k-1}) + \sum_{i=2}^k \mathcal{B}(k,i,0.5)\mathcal{X}_{i} ~,
\label{eqn:thm_tree2}
\enq
\beq
J_k ~=~  \mathcal{B}(k,0,0.5)J_k + \mathcal{B}(k,1,0.5)(1+J_{k-1}) +
\sum_{i=2}^k \mathcal{B}(k,i,0.5)J_{i} ~,
\label{eqn:thm_tree3}
\enq
for $k=2,3,\cdots$, with $\mathcal{X}_0 = \mathcal{X}_1 = 1$ and $J_0 =0$, $J_1 = 1$.
\end{proposition}
\begin{proof}
Noticing that the CR epoch termination event is a renewal event
under the fully-loaded assumption, the result can be easily
derived by extending the recursion analysis in \cite{MP:93} and using
the renewal-reward theorem \cite{G:96}. The details are omitted
due to space limitation.
\end{proof}
Since the GTA protocol is inspired by the simplified collision
model, the main idea is to assign a single slot exclusively for
transmission of each colliding user (that was not pruned by
the algorithm). The resulting DMT, therefore, is given in terms of
a single-user performance, i.e., $d_1^{MAC}(.)$. The main drawback
of the algorithm is the relatively large number of slots needed to
resolve each collision, which translates into a loss in the
effective multiplexing gain, i.e., the argument of $d_1^{MAC}(.)$
in (\ref{eqn:thm_tree1}). It is now easy to see that GTA
cannot achieve the full effective multiplexing gain of
the multiple access channel, i.e., $\min\{KM,N\}$. An example
highlighting this fact will be provided in the later part of this
section. On the other hand, the DMT in \eqref{eqn:thm_tree1}
reveals the performance dependence on $p_t$ (and $r$), which
implies the possibility of maximizing the diversity gain by
choosing the appropriate values, $p_t^*$ and $r^*$ for each $r_e
\in [0, \min\{KM,{N}\})$. At the moment, we do not have a general
analytical solution for this optimization problem. However, the
solution for the special case of two users is obtained in
Section~\ref{dmdt-examples}.

Next, we characterize the optimal DMT for the O-NDMA protocol
(Again we do not have a delay parameter in the tradeoff since the
number of ARQ rounds is {\bf always} equal to the number of
colliding users).
\begin{proposition} \label{thm:dmdt_ndma}
The \emph{optimal} DMT for O-NDMA is, \beq d^{ONDMA}(r_e) ~=~
d_1^{MAC}\left(r_e\right). \label{eqn:ondma_d_opt} \enq
\end{proposition}
\begin{proof}
The DMT for O-NDMA with a given $p_t \in (0,1]$ and $r$ is found as
\beq \label{eqn:dmt_ndma} d^{ONDMA}(r_e)
~=~ d_1^{MAC}(r) \quad \mbox{where} \quad r ~=~\frac{Kp_t+(1-p_t)^K}{Kp_t}r_e, 
\enq utilizing the average throughput results in \cite{TZB:00}, and 
noting that the average SNR of each single-user decoder is still $\rho$.
Then, it is easy to find that the optimal values $(r^*,p_t^*) =
(r_e,1)$, which yields \eqref{eqn:ondma_d_opt}. We omit the details
due to space limitation.
\end{proof}
The matched-filter-like structure utilizing the orthogonality of
transmissions over different slots allows the O-NDMA protocol to
achieve the single-user performance, as we see from
\eqref{eqn:dmt_ndma}. Furthermore, $p_t^*=1$ ensures that the
throughput is maximized, and the optimal DMT is given by
\eqref{eqn:ondma_d_opt}. By comparing the expressions in
\eqref{eqn:thm_tree1} and \eqref{eqn:ondma_d_opt}, we realize that
the O-NDMA protocol achieves a larger diversity gain, as compared
with the GTA protocol, for any $r_e$ less than $\min\{M,N\}$.

Finally, the optimal DMDT of the IR-ARQ random access protocol is
characterized in the following theorem.
\begin{theorem} \label{thm:dmdt}
The \emph{optimal} DMDT for the IR-ARQ protocol is, \beq
\label{eqn:d_opt} d^{IR}(r_e,L) ~=~
d_K^{MAC}\left(\frac{r_e}{KL}\right). \enq
\end{theorem}
\begin{proof}
(sketch) First, we assume an asymptotically large block length
$T\rightarrow\infty$ to allow our error correction (and detection)
scheme to operate arbitrarily close to the channel fundamental
limits. An application of the renewal-reward theorem \cite{G:96}
gives \beq \label{eqn:tput} \eta_{FL} ~=~
\frac{p_tKR}{1+\sum_{k=1}^K \mathcal{B}(K,k,p_t) \sum_{
\ell=1}^{L-1} \beta_k(\ell)} ~. \enq In addition, given that joint
typical-set decoders \cite{CT:01}, which have an inherent ability 
to detect errors, are used for the channel output over slots 
$1$ to $\ell$ in ARQ round $\ell$,
extending the results in \cite{ECD:04} and \cite{TVZ:04}, the
probability of error $P_e$ is upper-bounded by, \beq
\label{eqn:petmp} P_e ~\le~ \sum_{k=1}^K \mathcal{B}(K,k,p_t)
\beta_k(L). \enq Noting that in the high SNR regime
$\beta_k(\ell)$ approaches \beq \label{eqn:beta1} \lim_{\rho \to
\infty} ~\beta_k(\rho, \ell) ~=~ \mathbf{1}\left(r
>\min\left\{\ell M, \frac{\ell N}{k}\right\}\right) ~\triangleq~
\left\{
\begin{array}{ll} 0, & r < \min\{\ell M, \frac{\ell N}{k}\}
\\ 1, & r > \min\{\ell M,\frac{\ell N}{k}\}
\end{array} \right., \enq
we find the DMDT with a given $p_t \in (0, 1]$ as \beq
\label{eqn:dmdt} d^{IR}(r_e,L) ~=~ d_K^{MAC}
\left(\frac{r}{L}\right), \enq where $r$ can be obtained from
$r_e$ using the relation (for $0 \le r \le \min\{M,N\}$), \beq
\label{eqn:r_e} r_e ~=~ \frac{p_t K r} {1 + \sum_{k=1}^K
\mathcal{B}(K,k,p_t) \sum_{l=1}^{ L-1} \mathbf{1}\left( r >
\min\{lM, \frac{lN}{k} \}\right)} ~, \enq from 
(\ref{eqn:tput}--\ref{eqn:beta1}), and the results in \cite{ECD:04,TVZ:04}.
Finally, we find the optimal values $(r^*,p_t^*)=(\frac{r_e}{K},
1)$, which gives \eqref{eqn:d_opt}. A detailed proof is provided
in Appendix \ref{app:dmdt}.
\end{proof}
Two remarks are now in order.
First, we elaborate on the intuitive justification for the
optimal values $(r^*,p_t^*)=(\frac{r_e}{K},1)$ for the IR-ARQ
protocol. In the asymptotic case with $\rho \rightarrow \infty$, 
the error probability is dominated by the worst case $K$-user
collision for any $p_t \in (0,1]$, which does not depend on $\rho$ 
by definition. This implies that choosing $p_t=1$ will maximize
the average throughput, without penalizing the asymptotic behavior
of the error probability. Now with $p_t=1$, choosing
$r^*=\frac{r_e}{K}<\min\{M,\frac{N}{K}\}$ will result in an
effective multiplexing gain equal to $r_e$ and will minimize the
number of rounds needed to decode the colliding messages, since
each user is transmitting at a small rate. Furthermore, it is clear
that with this choice of $r^*$, we can achieve any desired
effective multiplexing gain less than $\min\{KM,N\}$ (the degrees
of freedom in the coordinated multiple access channel).
Next, Comparing Propositions \ref{thm:dmdt_gta},
\ref{thm:dmdt_ndma} and Theorem \ref{thm:dmdt}, it is
straightforward to verify that the DMT of the IR-ARQ protocol is
{\bf always} superior to that of the GTA and O-NDMA protocols.
This advantage of IR-ARQ is a manifestation of the {\bf ARQ
diversity} resulting from the IR transmission and joint decoding.
More specifically, the ARQ diversity {\bf scales down} the
effective multiplexing in the right hand side of
(\ref{eqn:d_opt}), and hence, results in an increased diversity
advantage (since $d_K^{MAC}(.)$ is a decreasing function in its
argument). The O-NDMA protocol does not allow for {\bf efficiently} 
exploiting the ARQ diversity due to the sub-optimality of
repetition based ARQ.

\subsection{Examples}\label{dmdt-examples}
We numerically illustrate the gains offered by the IR-ARQ protocol,
as compared with the GTA and O-NDMA protocols, in the following
two examples.

\subsubsection{Two-User Scalar Random Access Channels}
We consider the single-antenna $2$-user random access channel,
i.e., $M=N=1$ and $K=2$. Substituting these parameters in
Proposition \ref{thm:dmdt_gta}, we obtain the DMT for the GTA
protocol as, $d^{GTA}(r_e) ~=~ d_1^{MAC}\left(
\frac{1+3p_t^2}{2p_t}r_e \right) ~=~ 1-\left(\frac{1+3p_t^2}
{2p_t}\right)r_e $.  In order to maximize the effective
multiplexing gain that achieves nonzero diversity gain, we need to
choose $p_t = \frac{1}{\sqrt{3}}$, which yields the optimal
DMT for GTA as $d^{GTA}(r_e) = 1-\sqrt{3}r_e, \ 0 \le r_e <
\frac{1}{\sqrt{3}}$. The optimal DMTs for O-NDMA and IR-ARQ
are obtained from Proposition \ref{thm:dmdt_ndma} and
Theorem \ref{thm:dmdt}. \figref{DMDT_Comp} compares the tradeoffs
of the three protocols where the IR-ARQ protocol is shown to
dominate our two benchmarks, with both $L=1, 2$. Even though
O-NDMA achieves the nominal single-user DMT {\bf without
multi-user interference}, i.e., $d(r_e)=1-r_e, \forall r_e < 1$,
it is still worse than IR-ARQ, since it wastes slots to facilitate
single-user decoding and relies on the repetition ARQ. In
addition, as $L$ increases from 1 to 2, the DMDT of IR-ARQ
improves, as expected.

\subsubsection{Two-User Vector Random Access Channels}
We consider a $2$-user vector random access channel with $M=1$ and
$N=2$. By allowing multiple antennas at the BS, the total degrees
of freedom of the system is increased by a factor of $2$, as
compared with the scalar channel in the previous example. The
tradeoffs achieved by the three protocols in this scenario are
shown in \figref{DMDT_Comp2}. First, we observe that the three
protocols achieve an increased diversity gain, for a given $r_e$,
when compared with the scalar channel in \figref{DMDT_Comp}.
However, the full effective multiplexing gain, $r_e=2$, is not
achieved by the GTA and O-NDMA protocols, since these two
protocols exclude the possibility of first-round decoding when a
collision occurs. The IR-ARQ protocol, on the other hand, achieves
$r_e=2$, and the DMDT further improves as $L$ increases.

\section{Random Arrivals} \label{sec:random}

In this section, we relax the ``fully loaded'' assumption
adopted in Section \ref{sec:dmdt}. In addition to the traditional
measures of stability region and average delay commonly used in
this set-up, we also consider the probability of error. In
particular, for the proposed IR-ARQ protocol, we will show,
through numerical results, that the choice of the
transmission-delay constraint $L$ determines an
\textbf{interesting tradeoff} between the average delay and error
probability: For typical SNR, a larger $L$ leads to an increase in
the average delay along with a decrease in the error probability.

We consider infinite-length queues at the users, that are fed by
randomly-arriving packets of a fixed length of $B_A$ information
bits. For simplicity, we assume that $B_T = B_A = B$, i.e., the
arrival packet size and the transmission packet size are the same.
Thus the first-round transmission rate $R$ is equal to the arrival
rate $R_A = (B/T)$. To emphasize that $R$ is a system parameter
determined by the arrival packet size, we denote the first-round
multiplexing gain $r$ by $r_A$ in this section, and call it
\emph{the arrival multiplexing gain}. The {\bf packet arrival}
rate of user $i$ is $\lambda_i = \lambda/K$ packets/slot, where
$\lambda$ denotes the total packet arrival rate, where arrivals
are assumed to be independent across users.

\subsection{Stability and Average Delay}
We use the following notion of stability~\cite{DST:03}: let
$\mathbf{g}(m) \triangleq (g_1(m), \cdots, g_K(m))$ be the vector
of the backlogs at the beginning of CR epoch $m$. Then, queue $i$
of the system is stable if $\lim_{m \to \infty} ~ \Prob \left(
g_i(m) < \bar{g} \right) = F(\bar{g})$ and $\lim_{\bar{g} \to
\infty} ~F(\bar{g}) = 1$. Furthermore, we say that the system is
stable if all the $K$ queues in the system are stable.

The stability region of GTA and O-NDMA can be found using the
techniques in \cite{MP:93,DST:03} as 
\beq \lambda
~<~ \frac{\sum_{k=0}^K \mathcal{B}(K,k,p_t) J_k} {\sum_{k=0}^K
\mathcal{B}(K,k,p_t) \mathcal{X}_k} ~, \qquad \mbox{and} \qquad
\lambda ~<~
\frac{Kp_t}{Kp_t+(1-p_t)^K}. \enq From the literature
(\cite{NMT:05} and references therein), we find only limited
analytical results on the average delay of slotted ALOHA channels.
In this paper, we present only numerical results for the average
delay of the GTA and O-NDMA protocols, and provide an approximate
delay analysis for the proposed IR-ARQ protocol. The average delay
of the IR-ARQ scheme can be approximated by using the analysis of
the M/G/1 queue with vacations \cite{BG:92}, following the
approach of Tsatsanis {\em et al.}~\cite{TZB:00}. This analysis
yields only an approximation of the average delay, since the CR
epoch lengths of the IR-ARQ scheme are dependent on the traffic
load (and hence are not independent and identically distributed
(i.i.d.) as needed for the result to hold). However, as we will
see, the i.i.d. property holds in the limit of $\rho \rightarrow
\infty$, and hence, our result becomes asymptotically accurate. We
also note that as $K$ increases, this approximation becomes
progressively more accurate \cite{TZB:00}. We summarize our
results in the following theorem.
\begin{theorem} \label{thm:stab}
Assuming that $\exists ~ \ell < \infty \mbox{ with } \ell \in \{1,
\cdots, L\}, \mbox{ such that } \alpha_K(\ell)>0$, the necessary
and sufficient condition for the stability of the IR-ARQ protocol
is (in packets/slot) \beq \label{eqn:stab_dom} \lambda ~<~
\frac{\eta_{FL}}{R} ~=~ \frac{p_tK}{1+\sum_{k=1}^K
\mathcal{B}(K,k,p_t) \sum_{\ell=1}^{L-1} \beta_k(\ell)} ~. \enq
For Poisson arrivals, when $\lambda$ satisfies
\eqref{eqn:stab_dom}, the average delay is \emph{approximately}
given by (in slots) \beq  \label{eqn:delay} D ~\approx~ \ex[U] +
\left(\frac{1}{p_t}-1\right) \ex[V] +\frac{\lambda \left( \ex[U^2]
+ \frac{(2-p_t)(1-p_t)}{p_t^2} \ex[V^2] + 2\left( \frac{1}{p_t} -1
\right) \ex[U] \ex[V] \right)}{2\left( K- \lambda \left( \ex[U]+
 \left( \frac{1}{p_t}-1 \right) \ex[V] \right) \right)} + \frac{\ex[V^2]}
{2\ex[V]} ~,
\enq
where the expected values are evaluated as,
\small
\beq
\ex [U] = 1 + \sum_{k=1}^{K} \mathcal{B}(K-1,k-1,p) \sum_{\ell=1}^{L-1}
\beta_k(\ell) \quad ; \quad  \ex [U^2] = 1 + \sum_{k=1}^{K}
\mathcal{B}(K-1,k-1,p) \sum_{\ell=1}^{L-1} (2\ell +1)\beta_k(\ell),\no
\enq
\beq
\ex[V] = 1 + \sum_{k=1}^{K-1} \mathcal{B}(K-1,k,p) \sum_{\ell=1}^{L-1}
\beta_k(\ell) \quad ; \quad \ex[V^2] = 1 +\sum_{k=1}^{K-1}
\mathcal{B}(K-1,k,p) \sum_{\ell =1}^{L-1} (2\ell +1)\beta_k(\ell), \no
\enq
\normalsize
\beq \label{eqn:p_eq} 
\begin{array}{lc} \mbox{ and $p \in (0,1]$ satisfies } \qquad \qquad 
& Kp ~=~
\lambda \left[ 1 + \sum_{k=1}^K \mathcal{B}(K,k,p)
\sum_{\ell=1}^{L-1} \beta_k (\ell) \right]. \end{array} \enq
Moreover, the
delay expression in \eqref{eqn:delay} holds with probability 1 if
$U$ and $V$ are i.i.d. and $U$ and $V$ are mutually independent.
\end{theorem}
\begin{proof}
See Appendix \ref{app:stab}.
\end{proof}
A few remarks are now in order:
First, the assumption in Theorem \ref{thm:stab} always holds when
$L$ is finite since the length of any CR epoch is bounded by $L$.
If $L = \infty$, this assumption requires the existence of a
non-zero probability that the length of an epoch is finite.
Second, as $\rho\to\infty$, the stability region
\eqref{eqn:stab_dom} approaches  \beq \label{eqn:stab_hsnr}
\lambda ~<~ \frac{p_tK}{1+\sum_{k=1}^K \mathcal{B}(K,k,p_t)
\sum_{\ell =1}^{L-1} \mathbf{1}\left(r_A>\min\left\{\ell
M,\frac{\ell N}{k} \right\}\right)} . \enq To achieve the maximum
stability region, we need to maximize the right hand side of
\eqref{eqn:stab_hsnr} over $p_t$. At the moment, we do not have a
general solution for this problem. Thus, we present results only
for one special case: $r_A<\min\{M, \frac{N}{K}\}$. In this case,
the stability region is $\lambda < p_t K$, and the maximum
stability region is thus given by $\lambda < K$ for the optimal
choice of $p_t = 1$. This is a remarkable improvement over the
O-NDMA protocol, whose maximum stability region is only $\lambda <
1$, for any $r_A$.
Finally, the diversity gain with random arrivals can be readily
obtained from the results in the previous section. The only
difference is that, unlike the fully-loaded case, one cannot
optimize over $r_A$ in this scenario. In summary, we find that the
GTA, O-NDMA and IR-ARQ protocols achieve the diversity gains 
$d^{GTA}(r_A) ~=~ d^{ONDMA}(r_A) ~=~ d_1^{MAC}(r_A)$ and 
$d^{IR}(r_A) ~=~ d_K^{MAC}\left(\frac{r_A}{L}\right)$, respectively. 

\subsection{Examples}
\subsubsection{Two-User Scalar Random Access Channels}
Here, we consider the random access channels with $M=N=1$. For ease
of analysis, we assume that $L \ge K=2$ for the IR-ARQ scheme.

The stability region of the different random access protocols with
$\rho \rightarrow \infty$ is summarized in Table
\ref{tbl:stab_ex}. In addition, the error probability, diversity
gain and average delay are shown in \figref{num_snr_pe1},
\figref{DMT_Stab} and \figref{num_lambda_D1} respectively. Here,
the stability region and diversity gain of the three protocols,
and the average delay of the IR-ARQ protocol with $\rho \to
\infty$, are obtained analytically. However, the average delay of
the GTA and O-NDMA protocols, and the average delay of the IR-ARQ
scheme with $\rho < \infty$ are obtained through numerical
simulations. In these simulations, we use $R = r_A \log (1+\rho)$ 
with $r_A=0.45$, and $p_t=1$ for the IR-ARQ and the O-NDMA
protocols, while $p_t=\frac{1}{\sqrt{3}}$ for the GTA protocol. It
is assumed that the transmission results in errors, if and only if
the channel is in outage~\cite{CT:01}; which is a valid assumption
if $T$ is sufficiently large. In addition, for the IR-ARQ
protocol, it is assumed that the errors in $\ell^{th}$ round,
where $\ell<L$, are always detected. We also note that when
$r_A<0.5$, the average delay expression for the IR-ARQ scheme,
evaluated from Theorem~\ref{thm:stab}, holds with probability $1$,
and is given by (when $p_t=1$) $D = 1.5 + \frac{\lambda}{2(2-\lambda)}$.
Table \ref{tbl:stab_ex} and \figref{DMT_Stab} shows that both the
stability region and diversity gain of the IR-ARQ protocol are the
largest. Next, we focus on the delay and the error probability of
IR-ARQ with different $L$'s and different $\rho$'s reported in
\figref{num_snr_pe1} and \figref{num_lambda_D1}. We observe that
the delay approaches the asymptotic result with $\rho=\infty$, and
the difference of the delay for IR-ARQ with $L=2$ and with $L=4$
decreases, as $\rho$ increases, which agrees with the analytical
results. Furthermore, \figref{num_snr_pe1} and
\figref{num_lambda_D1} reveal an important insight into the
relation between the performance of IR-ARQ and the
transmission-delay constraint $L$, i.e., a tradeoff between
average delay and error probability emerges. These figures suggest
that for certain finite $\rho$'s, a large $L$ achieves a small
error probability, at the expense of a large average delay and a
small stability region. Therefore, depending on quality-of-service
(QoS) requirements, $L$ can be adjusted for achieving the best
performance.

\subsubsection{Two-User Vector Random Access Channels}
Here, we consider the $2$-user random access protocols with $M=1$
and $N=2$ in the high SNR regime ($\rho \rightarrow \infty$). We
first note that the stability region and delay of the GTA and
O-NDMA protocols are not different from the scalar case; only the
diversity gain changes with this multiple-antenna setting. For the
IR-ARQ protocol, on the other hand, the average delay is given by
$D = 1.5 + \frac{\lambda}{2(2-\lambda)}$ for any $r_A \in [0,1)$, 
and the stability region
is given by, $\lambda < 2, \ \ 0 \le r_A < 1$, with $p_t=1$.
Comparing the stability region of the vector IR-ARQ protocol with
that of the scalar IR-ARQ protocol, we find that the vector IR-ARQ
achieves a better stability region, especially when $r_A
>0.5$. Finally, \figref{DMT_Stab2} shows the diversity gain
achieved with different random access protocols. As expected, the
IR-ARQ protocol achieves the best diversity gain, which improves as
$L$ increases.

\section{Conclusions}  \label{sec:conc}
We have proposed a new wireless random access protocol which
jointly considers the effects of collisions, multi-path fading,
and channel noise. The proposed protocol relies on incremental
redundancy transmission and joint decoding to resolve collisions
and combat multi-path fading. This approach represents a marked
departure from traditional collision resolution algorithms and exhibits
significant performance gains, as compared with two benchmarks
corresponding to the state of the art in random access protocols;
namely GTA and O-NDMA. It is interesting to observe
that, in order to fully exploit the benefits of the proposed
IR-ARQ protocol, all the users with non-empty queues must transmit
with probability one, when given the opportunity, and should use a
small transmission rate. Finally, we have identified the tradeoff
between average delay and error probability exhibited by the
IR-ARQ protocol for certain SNRs, and have shown that this
tradeoff can be controlled by adjusting the maximum number of ARQ
rounds.

\appendices

\section{Proof of Theorem~\ref{thm:dmdt}} \label{app:dmdt}
To find the long-term average throughput of IR-ARQ, we first focus on the
distribution and the expected value of the relevant and the irrelevant
epochs for user 1, $U$ and $V$. The probability mass functions (pmf) of
$U$ and $V$ are
\beqa
\Prob (U = \ell) &=&
\sum_{k=1}^{K} \mathcal{B}(K-1,k-1,p_t) \alpha_k(\ell), \quad (\ell=1, \cdots, L),
\label{eqn:P_U} \\
\Prob(V = \ell)
&=& \left\{ \begin{array}{ll}
\sum_{k=1}^{K-1} \mathcal{B}(K-1,k,p_t) \alpha_k(1) ~+~ (1-p_t)^{K-1}, &
\ell=1, \\ \sum_{k=1}^{K-1} \mathcal{B}(K-1,k,p_t) \alpha_k(\ell),
& \ell=2, \cdots, L \end{array} \right.
\label{eqn:P_V}
\enqa
We introduce the relation shown in \cite{ECD:04} for deriving the expected
values of $U$ and $V$:
\beq \label{eqn:alpha_beta}
\sum_{\ell=1}^L \left[ \sum_{i=1}^{\ell} a_i \right]\alpha_k(\ell) ~=~
\sum_{\ell=1}^L a_{\ell} \beta_k(\ell-1),
\enq
for any $(a_1, \cdots, a_L) \in \mathbb{R}^L$. Using this relation, it is
straightforward to get
\small
\beq
\ex[U] ~=~ \sum_{k=1}^{K} \mathcal{B}(K-1,k-1,p_t) \sum_{\ell=1}^{L}
\beta_k(\ell-1) \quad; \quad
\ex[V] ~=~ (1-p_t)^{K-1} + \sum_{k=1}^{K-1} \mathcal{B}(K-1,k,p_t)
\sum_{\ell=1}^{L} \beta_k(\ell-1).
\label{eqn:E_V}
\enq
\normalsize

Now, we are ready to calculate the long-term average throughput 
\eqref{eqn:tput} and the upper-bound of error probability $P_e$ given in
\eqref{eqn:petmp} for the IR-ARQ scheme as in the following.
First, we prove \eqref{eqn:tput} utilizing the renewal theory \cite{G:96}. Denoting the
average throughput of user 1 by $\eta_1$, the average throughput of the symmetric
system is given by $\eta_{FL} ~=~ K \eta_1 ~$.
Under the fully-loaded assumption, the event that a CR epoch terminates is a
renewal event.
We associate a random reward $\mathcal{R}$ to the occurrence of the renewal
event; $\mathcal{R} = R$ BPCU if the CR epoch is a relevant epoch for
user 1, and $\mathcal{R} = 0$ otherwise. Then, the renewal-reward theorem
\cite{G:96} with \eqref{eqn:E_V} gives,
\beq \label{eqn:eta}
\eta_1 ~=~ \lim_{s \to \infty} ~\frac{b_1(s)}{sT}
~=~ \frac{\ex[\mathcal{R}]}{\ex[\mathcal{X}]}
~=~ \frac{p_t \cdot R + (1-p_t) \cdot 0}{p_t \cdot \ex[U] + (1-p_t) \cdot \ex[V]}
~=~ \frac{p_t R}{1+\sum_{k=1}^K \mathcal{B}(K,k,p_t) \sum_{\ell=1}^{L-1} \beta_k(\ell)}.
\enq
Since $\eta_{FL} = K\eta_1$, we obtain $\eta_{FL}$ as given in \eqref{eqn:tput}.
Next, we prove \eqref{eqn:petmp}. An error occurs in
the IR-ARQ system in two different cases: (i) when decoding
failure is not detected at the BS and an ACK is fed back, or (ii)
when decoding fails at round $L$. Let $E_k(\ell)$ denote the event
that the decoder makes an error with $\ell$ received blocks when
$k$ users have collided in the first round. Then, we can
upper-bound $P_e$ as \cite{ECD:04} \beqa \no P_e &=& \sum_{k=1}^K
\mathcal{B}(K,k,p_t) \sum_{\ell=1}^L \Prob(
E_k(\ell),\mbox{an epoch length when $k$ users have collided}=\ell) \\
&\le & \sum_{k=1}^K \mathcal{B}(K,k,p_t) ~\Prob(E_k(L),\bar{\mathcal{A}_1},\cdots,
\bar{\mathcal{A}}_{L-1}) + \epsilon
~=~ \sum_{k=1}^K \mathcal{B}(K,k,p_t) \beta_k(L) + \epsilon, \label{eqn:p_e}
\enqa
where $\epsilon \to 0$ as $T \to \infty$. The intuition behind this upper-bound
is that the undetected error probability approaches zero as $T \to \infty$ for
the joint typical-set decoder, and hence the error probability is dominated by
the information outage probability. 

On the other hand, the diversity gain \eqref{eqn:dmdt} can be found as 
in the following. We first find
$\beta_k(\rho,\ell) ~\dot{=}~\rho^{-d_{k}^{MAC}(r/\ell)} $, 
using the results in \cite{ECD:04} and \cite{TVZ:04},
where $A(\rho) \dot{=} \rho^{-b}$ implies $b=-\lim_{\rho
\rightarrow \infty} \frac{\log_2 A(\rho)}{\log_2 \rho}$ as defined
in \cite{ZT:02} and $\dot{\le}$, $\dot{\ge}$ are similarly defined.
With this, given that $p_t$ does not depend on $\rho$, 
\eqref{eqn:petmp} implies that $P_e(\rho)$ satisfies 
the exponential inequality $P_e(\rho)
~\dot{\le}~\rho^{-d_K^{MAC}\left(\frac{r}{L}\right)}$, as $T \to
\infty$. In addition, applying the outage bound in \cite{ECD:04}
to the random-access system yields $P_e(\rho) ~\dot{\ge}~
\rho^{-d_K^{MAC}\left(\frac{r}{L}\right)}$. These two exponential
inequalities imply \eqref{eqn:dmdt}.

Noticing that the span of $d^{M,N}(r)$ is $r \in [0, \min\{M,N\})$ \cite{ZT:02},
we verify that the span of $d_{k}^{MAC}\left(\frac{r}{\ell}\right)$ is
$r \in [0, \min\{\ell M,\frac{\ell N} {k}\} )$.
Then \eqref{eqn:beta1} can be readily verified as in the following.
For $r < \min\{\ell M,\frac{\ell N} {k}\}$, it is obvious that
$\beta_k(\rho,\ell) \rightarrow 0$ as $\rho \rightarrow \infty$
from $\beta_k(\rho,\ell) ~\dot{=}~\rho^{-d_{k}^{MAC}(r/\ell)} $.
On the other hand, if $r>\min\{\ell M,\frac{\ell N} {k}\}$, then
$\beta_k(\rho,\ell) \rightarrow 1$ as $\rho \rightarrow \infty$
since the outage probability approaches $1$ as $\rho \to \infty$,
and the error probability given the outage event approaches $1$
as $T \to \infty$ by the strong converse \cite{G:68}.
Combining the results shown above, we get the DMDT of the proposed
IR-ARQ protocol as \eqref{eqn:dmdt}, where the relation between
$r$ and $r_e$ is given in \eqref{eqn:r_e}.

Finally, we consider the optimal pairs $(r^*,p_t^*)$ that achieve the
largest diversity gain for a desired $r_e$.
Regarding $r_e$ in \eqref{eqn:r_e} as a function of $r$, we observe that
$r_e(r)$ is discontinuous at the points $r=\min\{\ell M,\frac{\ell N}{k}\}$,
$\ell=1,\cdots,L-1$ and $k=1,\cdots,K$. We consider the values of $r$ that
lie within the first discontinuity of $r_e(r)$, i.e., $r \in [0, \min\{M,
\frac{N}{K}\})$. For these $r$, \eqref{eqn:r_e} yields
$r_e ~=~ p_t K r$.
Since the diversity gain increases when $r$ decreases, we want to determine
the smallest value of $r$ that achieves the desired $r_e$.
We find that $r$ is minimized when $p_t=1$. Thus for $r \in [0, \min\{M,
\frac{N}{K}\})$, the optimal choice of $(r,p_t)$ achieving $r_e$ is
$(\frac{r_e}{K},1)$. Furthermore, we find that
this choice achieves the maximum effective multiplexing gain given by the
degrees of freedom of the channel ($\min\{KM,N\}$). Thus we do not need
to consider the values of $r > \min\{M,\frac{N}{K}\}$, since it is clear
from \eqref{eqn:r_e} that such $r$ values result in a smaller diversity
gain for the same desired $r_e$.

\section{Proof of Theorem~\ref{thm:stab}} \label{app:stab}
We consider the backlog evolution $\mathbf{g}(m)$ of IR-ARQ,
where $m$ is the \emph{epoch} index. We observe that $\mathbf{g}(m)$
is an embedded Markov chain; $g_i(m)$, the backlog evolution of user
$i$ is,
\beq
g_i(m+1) = \left\{ \begin{array}{ll}
(g_i(m) - 1)^+ + a_i(m), & \mbox{ with probability } p_t \\
g_i(m) + a_i(m), & \mbox{ with probability } 1-p_t
\end{array}
\right.
\label{eqn:backlog}
\enq
where $a_i(m)$ is the number of packets that arrived at user $i$'s queue
during epoch $m$, and $(x)^+ = x$ if $x \ge 0$, $(x)^+ = 0$ otherwise,
for a real number $x$.

We first prove that \eqref{eqn:stab_dom} is the necessary and
sufficient condition for the stability of IR-ARQ. One can
straightforwardly prove that under the assumption that there is
a finite $\ell$ with $\alpha_K(\ell)>0$,
$\mathbf{g}(m)$ is a homogeneous, irreducible and aperiodic Markov chain,
by following the argument in the proof of Proposition 1 in \cite{DS:02}.
Given that the Markov chain has these properties, stability of the system is
equivalent to the existence of a limiting distribution for the Markov chain,
and thus is also equivalent to ergodicity of the Markov chain
\cite{AT:05,TM:79}. Sufficiency and necessity of \eqref{eqn:stab_dom} for
the ergodicity can be straightforwardly proved by following
the footsteps of the proof of Theorem 1 in \cite{AT:05}.
In particular, we consider a stochastically dominant system, in which users
are first chosen according to the probability-$p_t$ rule, and those users
with empty queues transmit {\em dummy} packets. It can be shown that
\eqref{eqn:stab_dom} is a sufficient condition for the stability of 
the dominant system, which implies the stability of the original system.
On the other hand, we observe that the bounded homogeneity
property \cite{TM:79} holds for the Markov chain \eqref{eqn:backlog}, 
and thus the instability of the dominant system implies the instability of
the original system.

Next, we consider the approximate average delay.
We denote the time period between the instance when a packet of user 1
reaches the head of its queue and the instance when it has finally been
transmitted, by a random variable $\mathcal{Y}$ (slots).
Then, the average delay $D$ is given by the result for the M/G/1 queue
with vacations \cite{BG:92},
$D = \mathbb{E}[\mathcal{Y}] +
\frac{\lambda \mathbb{E}[\mathcal{Y}^2]}{2(1-\lambda \mathbb{E}[\mathcal{Y}])}
+ \frac{\mathbb{E}[V^2]}{2\mathbb{E}[V]}$,
with probability 1, if $\mathcal{Y}$ is i.i.d. and $V$ is i.i.d..
Here, the first moment of $\mathcal{Y}$ is calculated as,
\beq
\no
\mathbb{E}[\mathcal{Y}] = \mathbb{E}[p_t U + p_t(1-p_t)(U+V)
+ p_t(1-p_t)^2(U+2V) + \cdots]
= \mathbb{E}[U] + \left(\frac{1}{p_t}-1\right)\mathbb{E}[V].
\enq
On the other hand, the second moment of $\mathcal{Y}$ is calculated as,
\beqa
\nonumber
\mathbb{E}[\mathcal{Y}^2] & = \mathbb{E}[p_t U^2 + p_t(1-p_t)(U+V)^2
+ p_t(1-p_t)^2(U+2V)^2 + \cdots] \\ \no
& = \mathbb{E}[U^2] + \frac{(2-p_t)(1-p_t)}{p_t^2}\mathbb{E}[V^2]
+ 2\left(\frac{1}{p_t}-1\right)\mathbb{E}[U]\mathbb{E}[V],
\enqa
if $U$ and $V$ are independent. Substituting the values of
$\mathbb{E}[\mathcal{Y}^2]$ 
and $\mathbb{E}[\mathcal{Y}]$ in $D$, the approximate delay
\eqref{eqn:delay} can be readily obtained.
The expected values of the steady-state epoch lengths,
$\mathbb{E}[U],\mathbb{E}[U^2],\mathbb{E}[V]$ and $\mathbb{E}[V^2]$
used in the expression \eqref{eqn:delay} are easy to derive utilizing
\eqref{eqn:alpha_beta}, noticing that the pmf's
for $U$ and $V$ are given by \eqref{eqn:P_U} and \eqref{eqn:P_V}
with a substitution of $p$ into $p_t$, where $p \triangleq p_t (1-p_e)$
and $p_e$ is the steady-state probability of a user's queue being empty.

Finally, to see that the steady-state transmission probability $p$
can be found by solving the equation \eqref{eqn:p_eq}, we consider the
method of the steady-state analysis of the Markov chain
$g_1(m)$, whose time-evolution is given by \eqref{eqn:backlog}.
Let us define the steady state values: 
$g_1 \triangleq \lim_{m \rightarrow \infty} g_1(m)$
and
$a_1 \triangleq \lim_{m \rightarrow \infty} a_1(m)$.
Then, taking expectation on both sides of \eqref{eqn:backlog}
results in:
\beq
\mathbb{E}[g_{1}(m+1)] = \mathbb{E}[g_{1}(m)] - p_t\Prob(g_{1}(m)>0)
+ \mathbb{E}[a_1(m)].
\enq
In the limit as $m \rightarrow \infty$, we have,
$\mathbb{E}[g_1]  = \mathbb{E}[g_1] - p_t\Prob(g_1>0) + \mathbb{E}[a_1]$,
or $p=\mathbb{E}[a_1]$. Thus,
\beq
p ~=~ \frac{\lambda}{K}
\left( p_t\Prob(g_1>0)\mathbb{E}[U]+(1-p_t\Prob(g_1>0))\mathbb{E}[V]
\right) ~=~ \frac{\lambda}{K}
\left( p\mathbb{E}[U]+(1-p)\mathbb{E}[V]
\right),
\label{eqn:p_empty2}
\enq
which is equivalent to \eqref{eqn:p_eq}.

\newpage

\begin{table*}[bp]
\caption{Stability Region of Different Two-User Scalar random access
Protocols}
\begin{center}
\begin{tabular}{|c|c|c|}
\hline
 & Stability Region & Maximum Stability Region \\
\hline \hline
GTA & $\lambda < 2p_t/(1+3p_t^2)$ & $\lambda < 1/\sqrt{3}$, with
$p_t=1/\sqrt{3}$ \\
\hline
O-NDMA & $\lambda < 2p_t/({2p_t+(1-p_t)^2})$ & $\lambda < 1$, with $p_t=1$ \\
\hline
IR-ARQ & $\left\{ \begin{array}{ll} \lambda < 2p_t, & r_A < 0.5, \\
\lambda < 2p_t/(1+p_t^2), & r_A > 0.5. \end{array} \right.$
& $\left\{ \begin{array}{ll} \lambda < 2, & r_A < 0.5, \\
\lambda < 1, & r_A > 0.5. \end{array} \right.$, with $p_t=1$. \\
\hline
\end{tabular}
\end{center}
\label{tbl:stab_ex}
\end{table*}

\putFrag{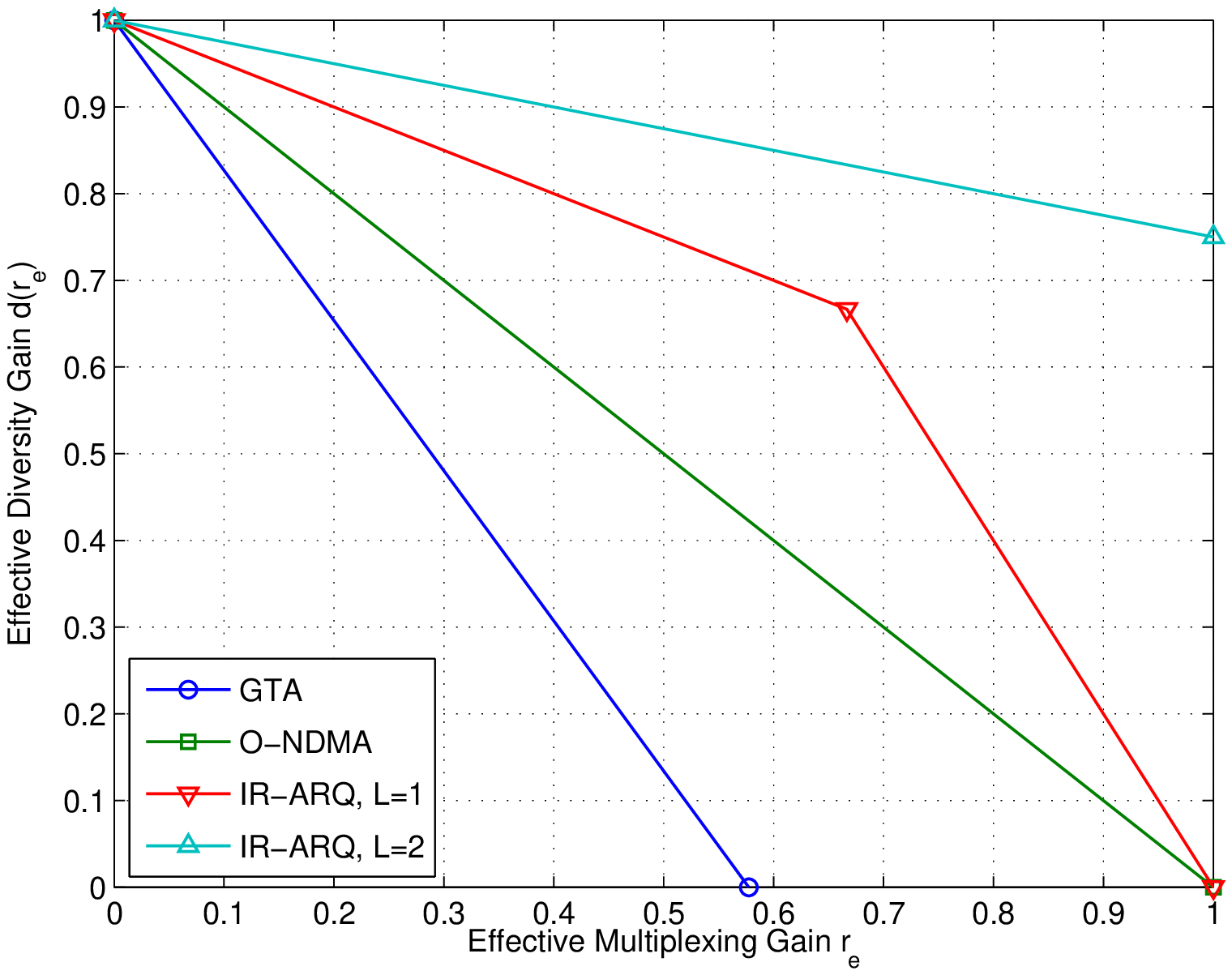}{Diversity-multiplexing tradeoff for various
two-user scalar random access systems.}{4.5}{}

\putFrag{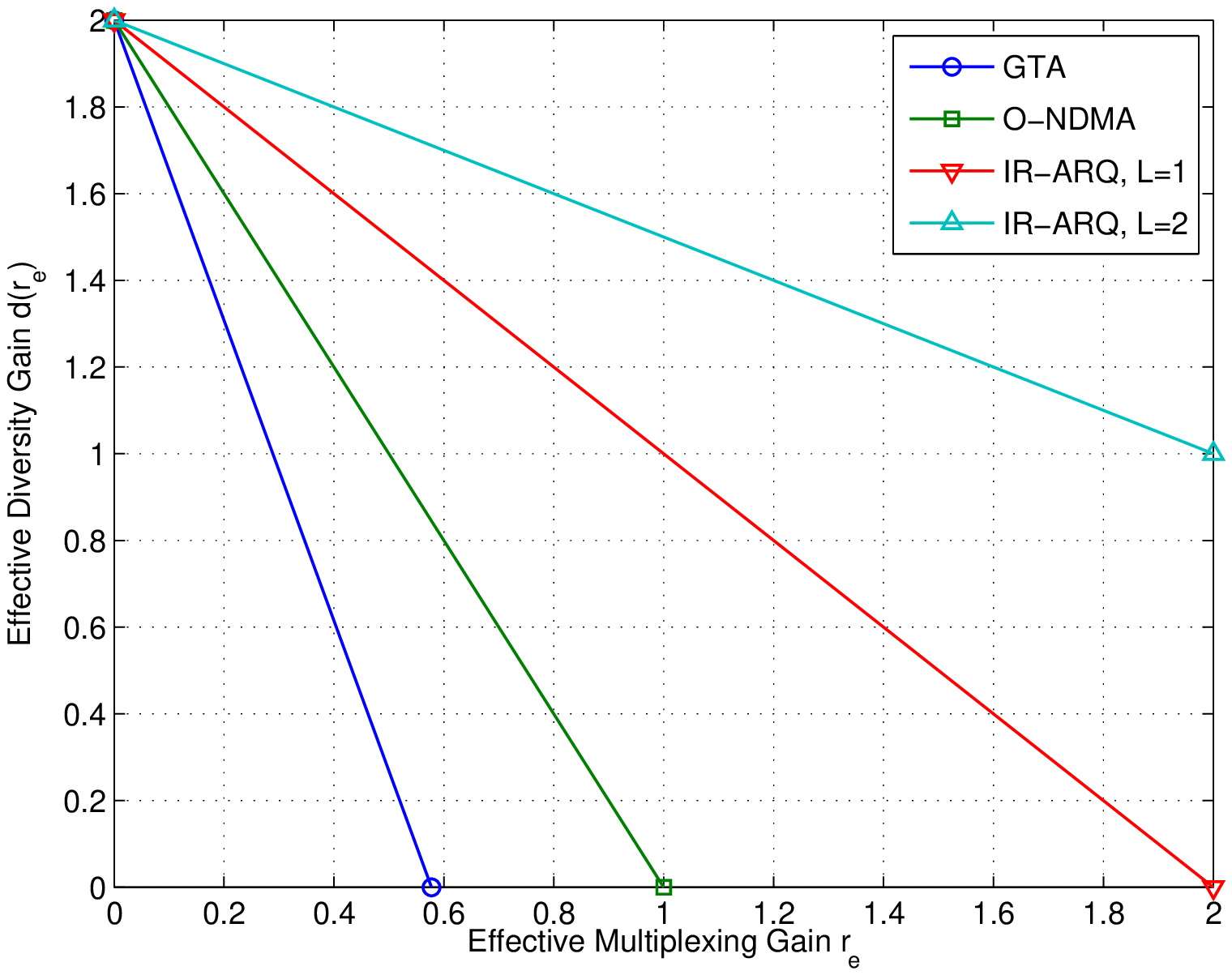}{Diversity-multiplexing tradeoff for various
two-user vector random access systems.}{4.5}{}

\putFrag{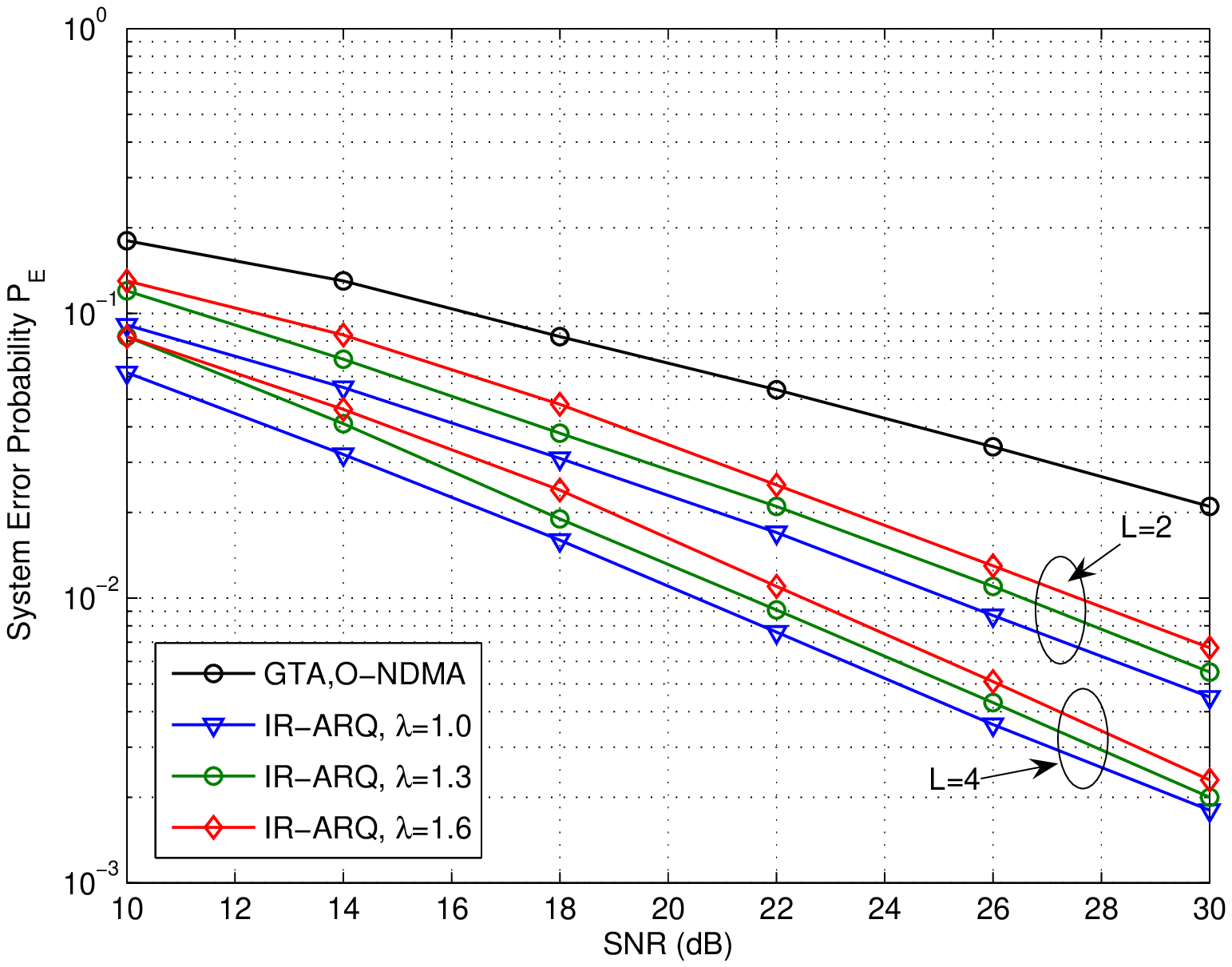}{System error probability versus SNR for various
two-user scalar random access systems with random arrivals. Here, $r_A=0.45$.}
{4.5}{}

\putFrag{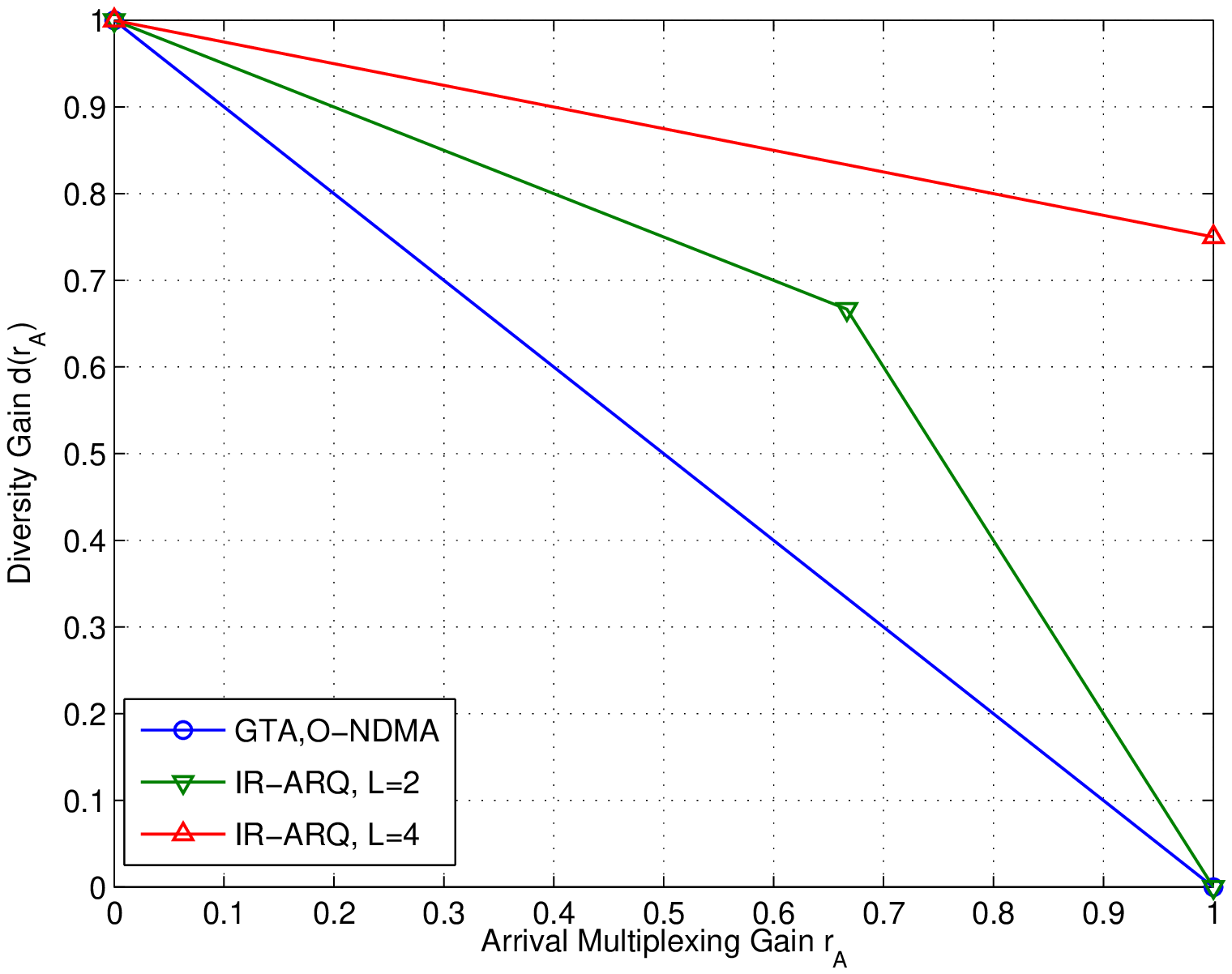}{Diversity gain versus the arrival multiplexing gain
$r_A$ for various two-user scalar random access systems with random
arrivals.}{4.5}{}

\putFrag{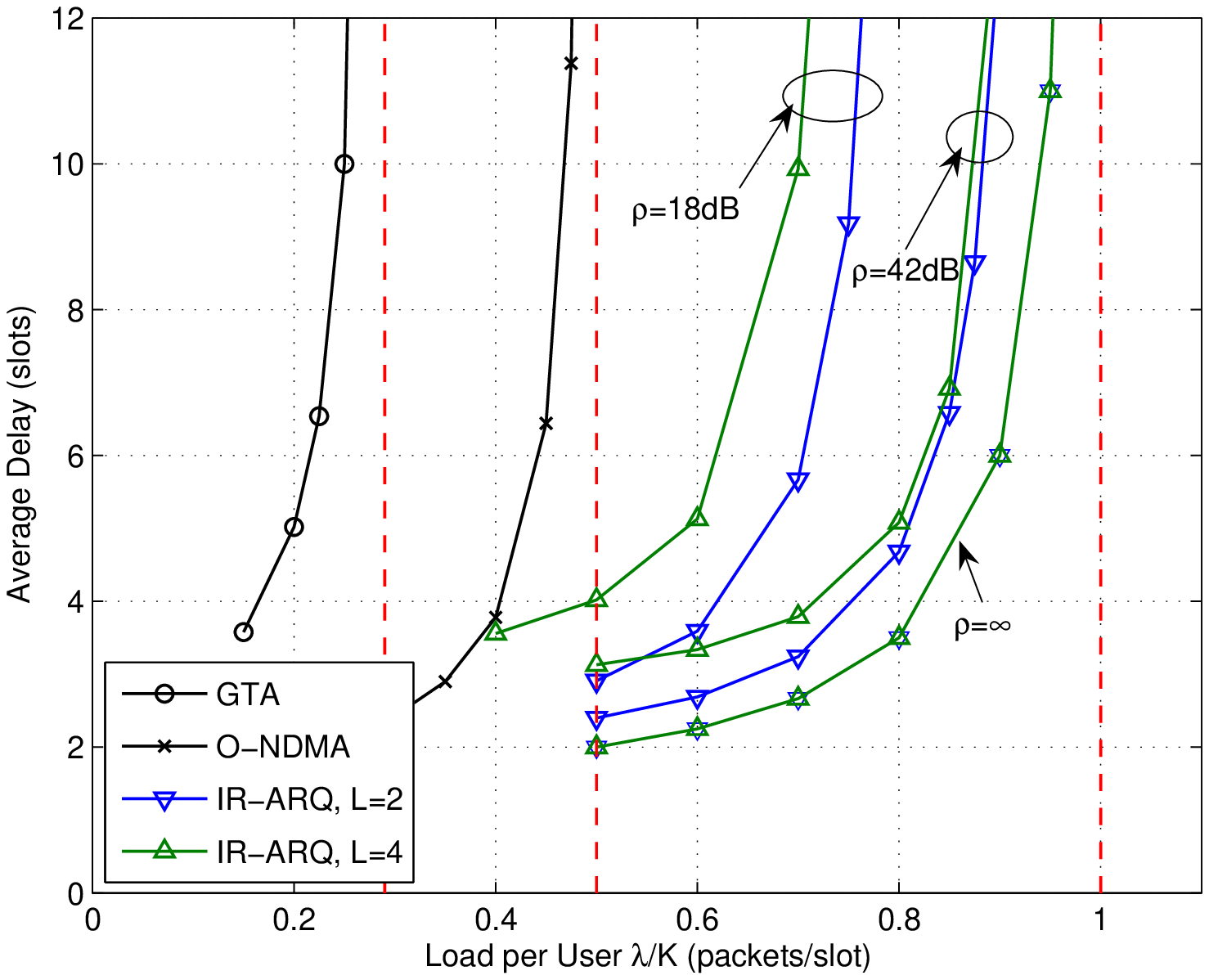}{Load per user versus average delay for various
two-user scalar random access systems with Poisson arrivals. Here, $r_A=0.45$.}
{4.5}{}

\putFrag{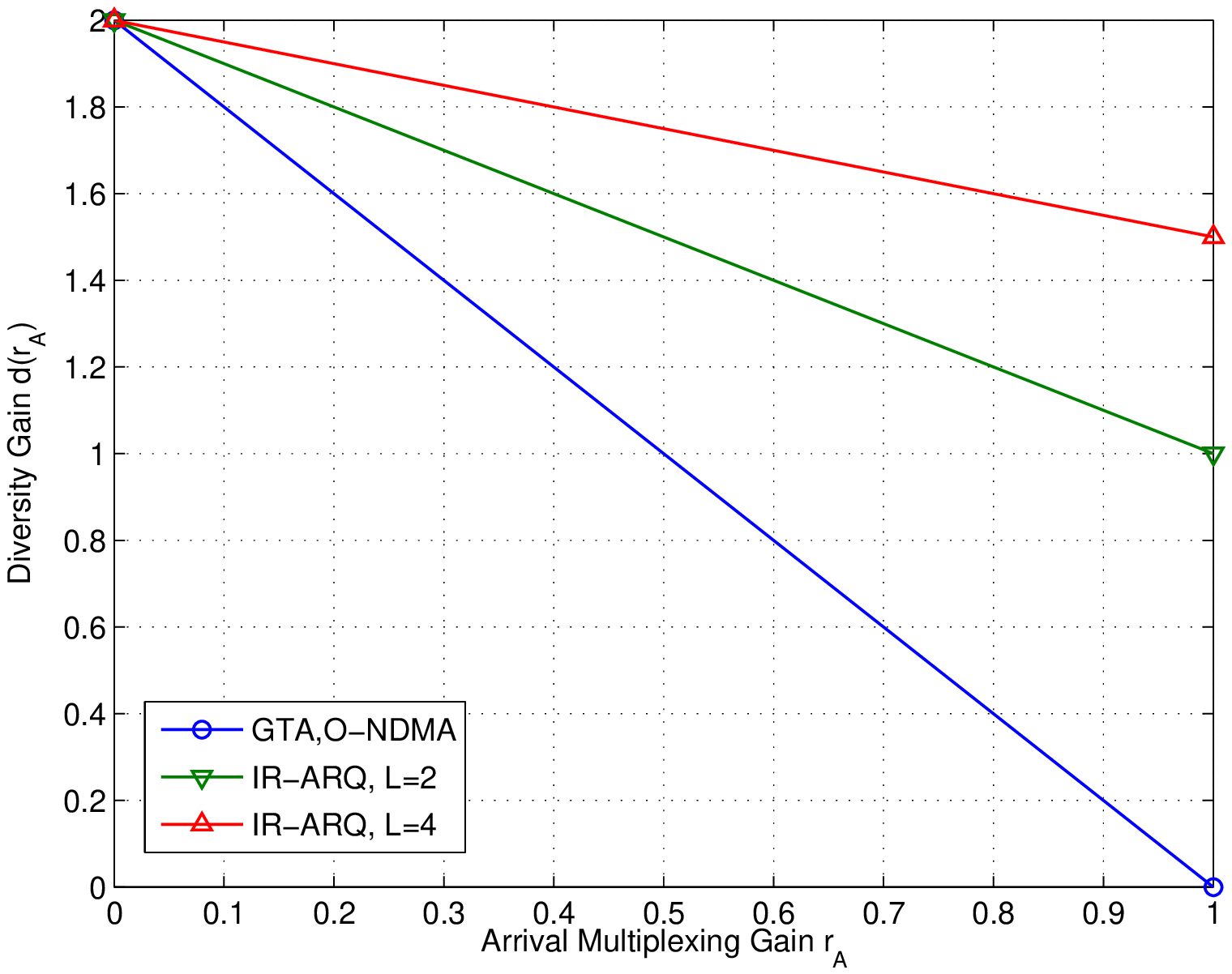}{Diversity gain versus the arrival multiplexing gain
$r_A$ for various two-user vector random access systems with random
arrivals.}{4.5}{}

\end{document}